\documentclass[11pt]{article}
\usepackage{a4wide,amssymb,graphics}

\newcommand{\be}{\begin{equation}}
\newcommand{\ee}{\end{equation}}
\newcommand{\bref}[1]{(\ref{#1})}
\newcommand{\sgn}{\, \mathrm{sgn}\,}
\newcommand{\diag}{\mathrm{diag}}
\newcommand{\rs}{r_\mathrm{s}}
\newcommand{\cn}{\, \mathrm{cn}\, }
\newcommand{\dn}{\, \mathrm{dn}\, }
\newcommand{\pt}{P_{\! \scriptscriptstyle T}}
\newcommand{\un}[1]{\, \mathrm{#1}}

\begin{document}

\title{\bf Brane World Linearized Cosmic String Gravity}

\author{S. C. Davis\thanks{E-mail: S.C.Davis@swansea.ac.uk} \\ 
\em Department of Physics, University of Wales Swansea, Singleton Park, \\
\em Swansea, SA2 8PP, Wales
}

\maketitle

\begin{abstract}
The gravitational properties of cosmic strings in the Randall-Sundrum
brane world scenario are investigated. Using a gauge in which the
brane remains straight, the leading order corrections to the metric on
the brane are determined. In contrast to their non-brane equivalents,
these strings have an attractive $1/r^2$ potential and a radially
dependent deficit angle. These two effects alter many cosmological
properties of the string, such as the formation of double images.
As a result of the attractive force the string will collect matter as it
moves through space. 
\end{abstract}

\hfill SWAT/269

\section{Introduction}

Cosmic strings are a type of topological defect formed at phase
transitions in some grand unified theories
(GUTs)~\cite{vilshel}. In the past their gravitational effects have
been considered as a possible explanation for the anisotropies in the
cosmic microwave background (CMB). Recent observations~\cite{CMBdata} and
simulations~\cite{strCMBsim} have raised doubts about this.

It has  been suggested that the physical universe may be a 3-brane
embedded in a higher dimensional spacetime~\cite{BWold}. Ordinary
matter is confined to this brane, while gravity is free to propagate
in the higher dimensional bulk. To be phenomenologically credible,
gravity on the brane must, at least at large distances, resemble
4-dimensional Einstein gravity. In a scenario proposed by Randall and
Sundrum the bulk space has a non-zero cosmological constant. This
gives a mass to the bulk gravitons, suppressing their effects. The
standard 3+1-dimensional gravity is produced by massless gravitons
trapped on the brane, with small corrections from the bulk. By tuning
the brane tension the effects of the bulk cosmological constant can be
cancelled on the brane.

In contrast to a point particle, a cosmic string (in the standard
4-dimensional universe) does not exert a force on nearby matter or
radiation. Its gravitational effects result from the the fact that the
space around it is approximately conical. If two initially parallel
particle paths pass either side of the string, the conical geometry
will cause their paths to converge. A related effect is the formation
of double images. The string's gravitational lensing means that if an
observer is on the opposite side of a cosmic string to a quasar, he
will see two images of it~\cite{vilen,gott}.

The conical geometry also produces a Doppler shift in light emitted
from objects moving relative to the  string (assuming the string is
between the object and the observer). When a string passes between an
observer and the cosmic microwave background this Doppler effect
will produce anisotropies~\cite{strCMB,gott}. It is the CMB anisotropies
predicted by this effect which have conflicted with observational
data, and so damaged the credibility of cosmic string models.

In this paper the gravitational field around a brane
world cosmic string is found, using a weak field approximation. The resulting
field is significantly different to that of a string in the standard
four dimensional universe. This means that any cosmological
properties of the string which arise from its gravity need to be
revised in the brane world scenario. This is in contrast to point-like
particles, where the brane world gravity is qualitatively the same as
in the standard case. The geodesics away from the string are also
determined. These are used to give a modified expression for the
separation of double images created by the string.

\section{Linearized Brane Gravity}

We will start by reviewing a formulation of linearized gravity in
which the brane is kept straight, even when matter is
present~\cite{lingrav}. The metric can be written as
\be
ds^2 = \hat g_{\mu \nu} dx^\mu dx^\nu + 2 n_\mu dx^\mu dy + (1+\phi) dy^2
\ee
with
\be
\hat g_{\mu \nu} = e^{-2|y|/\ell} (\eta_{\mu \nu} + h_{\mu \nu})
\ee
$x^\mu$, with $\mu =0,1,2,3$, are the coordinates on the brane and the
fifth coordinate $y$ is chosen so that the brane is at $y=0$.
$\hat g_{\mu \nu}$ gives the induced metric on the
hypersurfaces $y= $ constant. The length $\ell$ is related to the bulk
cosmological constant by $\Lambda = 6/\ell^2$. $\ell$ gives the
characteristic length
scale of the brane corrections to the usual 3+1-dimensional Einstein gravity.
I will use the four dimensional Minkowski metric
$\eta_{\mu \nu} = \diag(-1,1,1,1)$. 

In the linearized approximation $h_{\mu \nu}$, $n_\mu$ and $\phi$ are
taken to be small. In this approximation the gauge conditions for the
brane to remain at $y=0$ are
\be
\phi= -\frac{\ell}{4} \sgn y \, h_{,y}
\label{gauge1}
\ee
\be
n_\mu = -\frac{\ell}{8} \sgn y \, h_{,\mu}
\label{gauge2}
\ee
\be
\tilde h_{\mu \nu}{}^{,\mu} = 0
\label{gauge3}
\ee
where $h = \eta^{\mu \nu}h_{\mu \nu}$ and
$\tilde h_{\mu \nu} = h_{\mu \nu} - (1/4) \eta_{\mu \nu} h$
is the traceless part of $h_{\mu \nu}$.

The energy momentum tensor is
\be
T^{\mu \nu} = -\frac{3}{4 \pi \ell}\sqrt{1-\phi} \, \hat g^{\mu \nu} \delta(y) 
	+ t^{\mu \nu}\delta(y) \ , \ \ T^{5 \mu} = T^{55} = 0  
\ee
The first term is the background from the brane, and
$t^{\mu \nu}$ is the matter perturbation on the brane. The
resulting Einstein equations (to leading order) are
\be
\partial_y \left(e^{-2|y|/\ell} \partial_y \tilde h_{\mu \nu} \right)
- \frac{2}{\ell} \sgn y \, e^{-2|y|/\ell} \partial_y \tilde h_{\mu \nu} 
+ \partial^\sigma \partial_\sigma \tilde h_{\mu \nu} = 
G_5 \delta(y)\left[-16 \pi  \left(t_{\mu \nu} 
- \frac{1}{3}\eta_{\mu \nu}t \right) 
- \frac{\ell}{2} h_{,\mu \nu} \right]
\label{eineq}
\ee
where $t = \eta^{\mu \nu} t_{\mu \nu}$.
$G_5$ is the 5-dimensional Newton's constant. It is related to the
fundamental 5-dimensional Planck mass $M_5$ by $G_5 = M_5^{-3}$. The effective
4-dimensional Newton's constant is $G_4 = M_4^{-2} = \ell^{-1} G_5$. Gravity
experiments constrain $\ell$ to be less than $1\un{mm}$, and so 
$M_5 \gtrsim 10^5 \un{TeV}$.
 
Taking the divergence of \bref{eineq} and using
\bref{gauge3} gives the constraint
\be
\left. \partial^\sigma \partial_\sigma h \right|_{y=0} 
	= \frac{32 \pi}{3 \ell} t \ .
\label{trh}
\ee
In momentum space \bref{eineq} has the solution
\be
\tilde h_{\mu \nu}(p,y) = 8 \pi G_5\left[ t_{\mu \nu} 
- \frac{1}{3}\left( \eta_{\mu \nu} - \frac{p_\mu p_\nu}{p^2} \right) t \right]
e^{2|y|/\ell} \frac{K_2(e^{|y|/\ell} p \ell)}{p K_1(p \ell)}
\label{linsol}
\ee
for $p^2 > 0$.

\section{Cosmic String Gravity}

The cosmic string metric on the brane can now be found by substituting
the string energy-momentum tensor into \bref{linsol} and setting $y=0$.
For simplicity I will take the limit of zero string width. In this case
its energy momentum is~\cite{vilen}
\be
t_{\mu \nu} = \mu \delta^2(x) \diag(1,0,0,-1)
\label{Tstr}
\ee
thus $t = -2\mu \delta^2(x)$. $\mu$ is the energy per unit length of
the string. It is proportional to the square of the symmetry breaking scale
at which the string formed. Note that since the string is `made' of
Higgs and gauge fields, it is confined to the brane and does not
extend into the fifth, bulk dimension. Substituting \bref{Tstr} into
\bref{linsol} and inverting the Fourier transforms gives 
\be
\tilde h_{33}(x,0) = -\tilde h_{00}(x,0) =
-\frac{8 \pi}{3} G_5\mu \int \frac{K_2(p \ell)}{p K_1(p \ell)} 
e^{i{\bf p}\cdot {\bf x}} \frac{d^2p}{(2\pi)^2}
\label{th03}
\ee
\be
\tilde h_{ij}(x,0) = \frac{16 \pi}{3} G_5\mu \int
\left(\delta_{ij} - \frac{p_i p_j}{p^2} \right) \frac{K_2(p \ell)}{p K_1(p \ell)}
e^{i{\bf p}\cdot {\bf x}} \frac{d^2p}{(2\pi)^2}
\label{thij}
\ee
with $i,j=1,2$ and ${\bf p} = (p_1, p_2)$, ${\bf x} = (x_1, x_2)$. 
Changing to cylindrical polar coordinates, \bref{thij} implies
\be
\tilde h_{rr}(x,0) = \frac{16 \pi}{3} G_5\mu \int
\sin^2(\theta-\beta) \frac{K_2(p \ell)}{p K_1(p \ell)}
\left(\sum_n i^n J_n(pr)e^{in(\theta-\beta)}\right)
\frac{p dp d\beta}{(2\pi)^2}
\label{thrr}
\ee
where ${\bf p} = p(\cos \beta, \sin \beta)$ and
${\bf x} = r(\cos \theta, \sin \theta)$.
Evaluating the angular integral simplifies \bref{thrr} to
\be
\tilde h_{rr}(r,0) = \frac{4}{3} G_5\mu \int
\frac{K_2(p \ell)}{p K_1(p \ell)} [J_0(pr) + J_2(pr)] \, p dp
\label{thrrb}
\ee
Similarly
\be
\tilde h_{\theta\theta}(r,0) = r^2 \frac{4}{3} G_5\mu \int
\frac{K_2(p \ell)}{p K_1(p \ell)} [J_0(pr) - J_2(pr)] \, p dp
\label{thttb}
\ee
\be
\tilde h_{33}(r,0) = -\tilde h_{00}(r,0) =
-\frac{4}{3} G_5\mu \int \frac{K_2(p \ell)}{p K_1(p \ell)} J_0(pr) \, p dp
\label{th03b}
\ee
Using \bref{trh} gives
\be
h(r,0) = \frac{16}{3} G_5 \mu \int \frac{2}{p^2 \ell} J_0(pr) \, p dp
\label{h}
\ee
Combining this with \bref{th03b} and simplifying gives 
\be
h_{33}(r,0) = -h_{00}(r,0) =
-\frac{4}{3} G_5 \mu \int \frac{K_0(p \ell)}{p K_1(p \ell)} J_0(pr) \, p dp
\label{h03}
\ee
The large $r$ behaviour of $h_{\mu \nu}$ could now be found by
expanding the above integrands around $p=0$ and evaluating the
integrals. Doing this gives $h_{\mu \nu} \sim \ln r$ for large $r$,
suggesting that the weak field approximation breaks down. A similar
situation arises around the standard non-brane cosmic
strings~\cite{vilen}. In fact this problem is just the result of
a poor choice of coordinates.

A better behaved solution is obtained by introducing a new radial
coordinate $r'$ satisfying
\be
\left[1+\frac{4}{3} G_4\mu \int \frac{\ell K_2(p \ell)}{p K_1(p \ell)} 
[J_0(pr) - J_2(pr)] + \frac{2}{p^2} J_0(pr) \, p dp \right]r^2
= \left[ 1 - 8 G_4\mu \{1 + f_1(r'/\ell)\}\right]r'^2
\label{rp}
\ee
where $f_1$ is the solution of the inhomogeneous differential equation
\be
\partial_x[xf_1(x)] = \frac{1}{3}\int \frac{K_0(q)}{K_1(q)} x J_1(q x) \, q dq
\label{f1}
\ee
with boundary condition $xf_1(x)|_{x \to \infty}=0$. 

Combining (\ref{thrrb},\ref{thttb},\ref{h}) and \bref{h03} with
\bref{rp}, we find to leading order in $G_4 \mu$ that the metric is
\be
ds^2 = (-dt^2 + dz^2)[1- G_4 \mu f_2(r'/\ell)] 
+ dr'^2 + [1 - 8G_4\mu \{1+f_1(r'/\ell)\}]r'^2 d\theta^2
\label{strg1}
\ee
with
\be
f_2(x) = \frac{4}{3} \int \frac{K_0(q)}{q K_1(q)} J_0(qx) \, q dq
\label{f2}
\ee
The metric for a cosmic string in the standard four dimensional
universe is obtained from \bref{strg1} by replacing $f_1$ and $f_2$
with zero. The resulting geometry is conical. Thus, outside the string core,
spacetime is locally flat.

For small $q$  
\be
\frac{K_0(q)}{q K_1(q)} \approx - \ln(q/q_*) 
+ \frac{1}{2}\left(\ln^2(q/q_*)  - \ln(q/q_*) + \frac{1}{2}\right)q^2 
+ O(q^4\ln^2 q)
\ee
with $q_*=2e^{-\gamma}$, where $\gamma$ is Euler's constant. Using this
approximation \bref{f1} and \bref{f2} are solved by
\be
f_1(r/\ell) \sim - \frac{2\ell^2}{3 r^2} 
\left[ 1 - \frac{2\ell^2}{9 r^2}\{ 12 \ln(r/\ell) - 5\} + \cdots \right]
\label{f1a}
\ee
\be
f_2(r/\ell) \sim \frac{4\ell^2}{3 r^2} 
\left[1 - \frac{2\ell^2}{r^2}\{2\ln (r/\ell) - 1\} + \cdots \right]
\label{f2a}
\ee
For brevity I have dropped the prime on $r'$. The above asymptotic
expansions are valid for $r \gg \ell$.

To leading order in $G_4 \mu$ and $\ell/r$ the metric \bref{strg1} is thus
\be
ds^2 = (-dt^2 + dz^2)\left[1- \frac{4 G_4 \mu \ell^2}{3 r^2}\right] 
+ dr^2 + \left[1-8 G_4\mu + \frac{16 G_4 \mu \ell^2}{3 r^2}\right]r^2 d\theta^2
\label{strg2}
\ee
Thus space is approximately conical for $r \to \infty$, with the
same deficit angle as the standard 4-dimensional cosmic string
spacetime~\cite{vilen}. The brane corrections give the string a
small $1/r^3$ attractive force, and a deficit angle which increases with $r$.

The approximation \bref{Tstr} assumes that all length scales of
interest are far bigger than the string width, $\rs \sim \mu^{-1/2}$. 
The GUT energy scale must be less than the five dimensional Planck
scale, so it may be lower than in the standard four dimensional
universe. Unless $\mu$ is far lower than $M_5^2$ (which is not to
be expected for a GUT string), $\rs \ll \ell$. In this case the
solution \bref{strg1} will still be valid for small $r$. For 
$\rs \ll r \ll \ell$ series expansions of \bref{f1} and \bref{f2} can be found
using the asymptotic expansion
\be
\frac{K_0(q)}{q K_1(q)} \sim \frac{1}{q} - \frac{1}{2q^2} + O(q^{-3})
\ee
This implies
\be
f_1(r/\ell) \approx \frac{\ell}{3 r}\ln(r/\ell) - \frac{1}{6} + \cdots
\label{f1b}
\ee
\be
f_2(r/\ell) \approx \frac{4\ell}{3 r} 
	\left[1+\frac{r}{2\ell}\ln(r/\ell) \right] + \cdots
\label{f2b}
\ee
For $r \sim G_4 \mu \ell = G_5 \mu$ the linearized gravity
approximation will break down. Unless $\mu \sim M_5^2$
this will occur inside the string core. Thus the above linearized string
gravity will usually be valid everywhere outside the string core. 
Note that as the string is approached, the gravitational
field does genuinely become stronger, in contrast to the previous $\ln r$
behaviour at large $r$ (see comments before eq.~\ref{rp}), which was
purely due to a poor choice of coordinates. This can be seen from the
behaviour of the Ricci tensor as $r \to 0$ (see appendix).

Far away from the string the gravitational field is very weak.
If $\ell \sim 1 \un{mm}$, then for a typical
GUT-scale string with $G_4 \mu \sim 10^{-6}$, the gravitational
acceleration at $r \sim 1 \un{pc}$ will be of order 
$10^{-42} \un{cm} \un{s}^{-2}$. Thus the string's
gravitational effects will not be significant at astronomical distances.

\section{Geodesics in a Cosmic String Background}

Using \bref{strg2} particle paths at large $r$ can be found. In
calculating the geodesic equations we must use the intrinsic
connection coefficients $\hat \Gamma^\mu{}_{\nu\lambda}$ (see
appendix), instead of the full 5-dimensional equivalents
$\Gamma^\mu{}_{\nu\lambda}$. This is because ordinary matter is
confined to the brane. If a particle were free to move in five
dimensions the usual geodesic equations (with
$\Gamma^\mu{}_{\nu\lambda}$) would apply.

To leading order in $G_4 \mu$, particle paths are given by
\be
\dot t = E[1+G_4 \mu f_2(r/\ell)]
\ , \ \ \ \ 
\dot z = P_z[1+G_4 \mu f_2(r/\ell)]
\ee
\be
\dot \theta = \frac{J}{r^2} [1+8G_4 \mu\{1+f_1(r/\ell)\}]
\label{thgeo}
\ee
\be
\dot r^2- G_4 \mu f_2(r/\ell) (\pt^2+m^2)
+ \frac{J^2}{r^2}[1+8G_4\mu \{1+f_1(r/\ell)\}] = \pt^2
\label{rgeo}
\ee
The dot denotes differentiation with respect to the geodesic's
parameter. If this parameter is taken to be $\tau/m$, where $\tau$ is
the proper time, then $E$, $J$, $P_z$ and $\pt$ are
respectively the energy and angular, $z$, and transverse momentum at
$r=\infty$. $m$ is the particle mass, so $\pt^2 = E^2 - P_z^2- m^2$. 

Combining \bref{thgeo} and \bref{rgeo}, and keeping only the leading
order (in $\ell/r$) terms of \bref{f1a} and \bref{f2a} produces 
\be
\left(\frac{\partial r}{\partial \theta}\right)^2 = 
\frac{\pt^2}{J^2}(1-16G_4 \mu) r^4 
- \left(1 - 8 G_4 \mu - \frac{4(9\pt^2+m^2) G_4 \mu \ell^2}{3J^2}\right) r^2
-\frac{16}{3}G_4 \mu \ell^2
\label{rthgeo}
\ee
The $O(G_4\mu)^2$ terms in the above expression have been dropped.

With the change of variables $w=\alpha/r$ and $\theta = \pm \beta u$, 
\bref{rthgeo} can be rewritten as 
\be
\frac{dw}{du} = -\sqrt{(1-w^2)(1 + k^2[w^2-1])}
\label{JEeq}
\ee
where $k$ is a suitably chosen constant.
This equation is solved by the Jacobian elliptic function
$w = \cn(u,k)$~\cite{grad}. Thus 
\be
r = \frac{\alpha}{\cn(\theta/\beta,k)} 
\label{rsol1}
\ee
The function $\cn(u,k)$ is periodic and can be thought of as a
generalisation of $\cos u$. It has period $4K$, where $K$ is the complete
elliptic integral $\int_0^{\pi/2} da (1-k^2 \sin^2 a)^{-1/2}$. The
function $1/\cn$ can be represented as a trigonometric
series involving terms proportional to $\sec \pi u/(2 K)$ 
and $\cos (2n-1)\pi u/(2 K)$ (for integer $n$). If $k$ is
small $K = (\pi/2)[1 + k^2/4 + O(k^4)]$ and 
\be
\frac{1}{\cn(u,k)} = \frac{\pi}{2K}\sec \frac{\pi}{2 K} u + O(k^2)
\label{cnapp}
\ee

To leading order in $G_4 \mu$
\be
\beta = 1+4G_4\mu + \frac{2(\pt^2+m^2)}{3 J^2}G_4 \mu\ell^2 + O(G_4 \mu)^2
\label{beta}
\ee
\be
k^2 = \frac{16 \pt^2}{J^2}G_4 \mu\ell^2 + O(G_4 \mu)^2
\label{ksq}
\ee
\be
\alpha = \frac{J}{\pt}\left(1 + 4G_4\mu 
- \frac{2(5\pt^2+m^2)}{3J^2}G_4\mu\ell^2\right)
\label{alpha}
\ee
Thus unless $\pt \ell \gg  J$, the parameter $k$ is
small. Substituting (\ref{cnapp}--\ref{ksq}) into \bref{rsol1} gives
the solution for $r$ (to leading order in $G_4 \mu$)
\be
r = \frac{J}{\pt} \sec \frac{\theta}{1+\epsilon}
\label{rsol2}
\ee
where
\be
\epsilon = 4G_4 \mu + \frac{2(3\pt^2+m^2)}{3J^2}G_4 \mu \ell^2 + O(G_4 \mu)^2
\label{epsdef}
\ee
This solution represents a particle approaching the string from
$\theta = \frac{\pi}{2}(1+\epsilon)$ and leaving at 
$\theta = -\frac{\pi}{2}(1+\epsilon)$. Thus the particle path is
deflected by an angle of $\pi \epsilon$. The corresponding result for
a non-brane string is obtained by setting $\ell = 0$.

If $k > 1$ \bref{JEeq} will instead be solved by $w = \dn(ku,k^{-1})$
(another Jacobian elliptic function). For small $k$, $\dn(u,k)$ is
approximately a constant plus small oscillatory terms proportional to
$\cos n\pi u/K$ (integer $n$). This solution thus represents a stable
orbit around the string. Normally a system with a $-1/r^2$ potential
would have orbiting solutions, but they would not be stable. However,
the spatially varying deficit angle of the string produces an
additional $1/r^4$ effective potential term in \bref{rgeo}, which
stabilises them. Note that these forces do not affect motion of
particles in the $z$-direction, thus this type of solution includes
helical as well as circular paths.

If $k$ is to be large, $J$ must be smaller than $\pt \ell \sqrt{G_4 \mu}$ 
[see \bref{ksq}]. In this case the $f_1$ term in \bref{rgeo} required to
stabilise the orbit is of similar size to terms that have been dropped
in the linearized gravity approximation. The stability of these orbits
may therefore disappear when higher order terms are considered. The
orbiting particles will then either escape to $r=\infty$ or fall into
the string core, where (since GUT processes are enhanced inside the
string) they are expected to decay into lighter fundamental
particles or string bound states. Whether the orbits are stable or
not, the string will collect matter as it passes through plasma, with
the trapped matter ending up in the string core or orbiting it.

\subsection{Double Images}

\begin{figure}
\centerline{\includegraphics{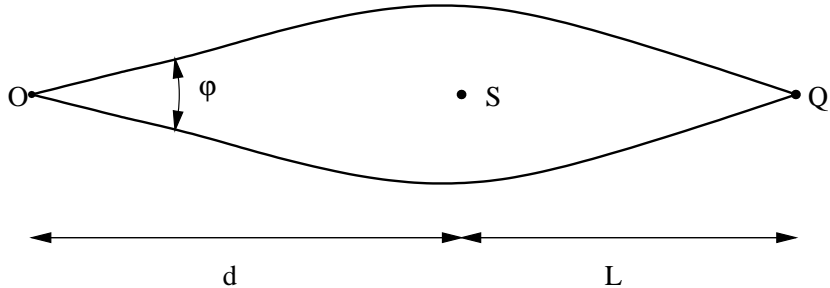}}
\caption{Double image of an object behind a string}
\label{fig:lb}
\end{figure}

One consequence of a cosmic string's gravitational field is the
formation of double images. This is illustrated in
figure~\ref{fig:lb}. Suppose a cosmic string (S) is positioned between
an observer (O) and a quasar (Q). Light rays passing either side of the
string will be bent towards it, and so rays from two different
directions can reach the observer. If the line OQ
makes an angle $\vartheta$ with the string, then the angular
separation of the two images is~\cite{vilen,gott}
\be
\varphi = \pi \epsilon \frac{L}{L+d} \sin \vartheta
\label{std2i}
\ee
where the assumptions $\varphi \ll 1$, $G_4 \mu \ll 1$ and 
$\varphi d\sin \vartheta, \varphi L \sin \vartheta \gg \ell$ have been used.
However, in contrast to the result for standard strings, the angle
$\pi \epsilon$ is dependent on $\varphi$. As $r \to \infty$, where
space is flat, the angular momentum satisfies $J = \pt \varphi d/2$. 
Substituting this into \bref{epsdef} gives an equation for
$\varphi$. For astronomical distances,  $\ell \ll d G_4 \mu$, and the
angle between the two images is approximately
\be
\varphi = \frac{8\pi L}{L+d}G_4 \mu \sin \vartheta
+ \frac{4\ell^2}{\pi d^2 G_4 \mu} \frac{L+d}{L}
\left(\sin \vartheta + \frac{m^2}{3(E^2-m^2)\sin\vartheta}\right)
\ee
The first term is the standard result which arises from the constant
part of the deficit angle. The second term gives the brane
corrections. For massless particles (e.g.\ photons) the final energy
dependent part is not present.

\section{Conclusions}

In this paper I have examined the linearized gravitational field around
a cosmic string in the Randall-Sundrum brane world scenario. The brane
corrections give the string an attractive $1/r^2$ potential, in sharp
contrast to standard cosmic strings. At astronomical distances the
force is very weak, and so significant gravitational effects
will only occur near the string core. Like the standard strings the
space at large radii is conical. However as the string core is
approached the deficit angle reduces.

Like their non-brane counterparts the brane strings produce double
images of objects behind them. The angle between these images will be
slightly larger than in the standard case. The closer the light rays
pass to the string, the more noticeable the difference. The brane
corrections will also alter the CMB anisotropies produced by the
string, although I will not consider this here. 

The attractive force of the brane string also means that particles can
make circular orbits around the string. If the particles also have
momentum in the direction parallel to the string, the corresponding
paths will be helical. The stability of these orbits is unclear, and
depends on gravitational effects beyond the linear
approximation. Whether the orbits are stable or not, the string is
expected to collect matter as it moves through the universe. This will
be in the form of particles which either orbit the string or are
pulled into its core. 

\section*{Acknowledgments}

I wish to thank Tim Hollowood and Warren Perkins for helpful
discussions, and PPARC for financial support.

\section*{Appendix}

For the metric \bref{strg2},
the connection coefficients intrinsic to the brane are
\be
\hat \Gamma^t{}_{tr} = \hat \Gamma^z{}_{zr} = 
\hat \Gamma^r{}_{tt} = - \hat \Gamma^r{}_{zz}
= -{1 \over 2}  G_4\mu \partial_r f_2(r/\ell)
\ee
\be
\hat \Gamma^r{}_{\theta\theta} = 
-\partial_r \left[{r^2 \over 2}(1- 8G_4\mu\{1+f_1(r/\ell)\}\right]
\ee
\be
\hat \Gamma^\theta{}_{r\theta} = \frac{1}{r} - 4G_4\mu \partial_r f_1(r/\ell)
\ee
The intrinsic Ricci tensor is
\be
\hat R_{tt} = -\hat R_{zz} = \frac{1}{2} G_4\mu 
\left[\frac{1}{r} \partial_r f_2(r/\ell) + \partial_r^2 f_2(r/\ell)\right] 
\ee
\be
\hat R_{rr} =  G_4\mu {1 \over r} \partial_r f_2(r/\ell)
\ee
\be
\hat R_{\theta\theta} = G_4\mu r^2 \partial_r f_2(r/\ell)
\ee 
Where the relationship $4xf_1''(x) + 8f_1'(x) + xf_2''(x) + f_2'(x) = 0$ 
has been used. The intrinsic curvature scalar on the brane is $\hat R = 0$.

Using \bref{f1b} and \bref{f2b} we see that as $r \to 0$,
$\hat R_{\mu\nu} \sim \ell/r^3$.

\providecommand{\pl}[3]{Phys.\ Lett.\ {\bf #1}, #3 (#2)}
\providecommand{\prd}[3]{Phys.\ Rev.\ D {\bf #1}, #3 (#2)}
\providecommand{\prl}[3]{Phys.\ Rev.\ Lett.\ {\bf #1}, #3 (#2)}
\providecommand{\nature}[3]{Nature {\bf #1}, #3 (#2)}
\providecommand{\apj}[3]{Ap.\ J. {\bf #1}, #3 (#2)}
\providecommand{\hepth}[1]{{\tt hep-th/#1}}

\end{document}